# Mapping Hydrogen Evolution Activity Trends of V-based A15 Superconducting Alloys

Peifeng Yu [a, 1], Jie Zhan [e, f, 1], Xiaobing Zhang [b], Kangwang Wang [a], Lingyong Zeng [a], Kuan Li [a], Chao Zhang [a], Longfu Li [a], Ying Liang [c], Kai Yan [d *], Yan Sun [e, f *], Huixia Luo [a *]

[a] School of Materials Science and Engineering, State Key Laboratory of Optoelectronic Materials and Technologies, Guangdong Provincial Key Laboratory of Magnetoelectric Physics and Devices, Key Lab of Polymer Composite & Functional Materials, Sun Yat-Sen University, Guangzhou 510275, China

[b] Beijing Advanced Innovation Center for Materials Genome Engineering, State Key Laboratory for Advanced Metals and Materials, University of Science and Technology Beijing, Beijing 100083, China

[c] Basic course department, Guangzhou Maritime University, Guangzhou 510725, China

[d] School of Environmental Science and Engineering, Sun Yat-Sen University, Guangzhou 510275, China

[e] Shenyang National Laboratory for Materials Science, Institute of Metal Research, Chinese Academy of Sciences, Shenyang, 110016, China

[f] School of Materials Science and Engineering, University of Science and Technology of China, Shenyang 110016, China

[1] These authors contributed equally

*Corresponding author:

Prof. Kai Yan (yank9@mail.sysu.edu.cn)

Prof. Yan Sun (sunyan@imr.ac.cn)

Prof. Huixia Luo (luohx7@mail.sysu.edu.cn)

**Abstract**

Exploring high-efficiency and low-cost electrocatalysts is valuable for water-splitting technologies. Recently, Si-group compounds have attracted increasing attention in electrocatalysis, considering the abundant Si-group elements on Earth. However, Si-group compounds for HER electrocatalysis have not been systematically studied. In this study, we unveil the activity trends of non-noble metal catalyst A15-type $V_3M$ (*i.e.*, $V_3Si$, $V_3Ge$, and $V_3Sn$) superconductors and show that $V_3Si$ is the most efficient HER catalyst because of the high electronic conductivity and suitable *d*-band center. Among them, the $V_3Si$ only requires 33.4 mV to reach 10 mA $cm^{-2}$, and only 57.6 mV and 114.6 mV are required to attain a high current density of 100 mA $cm^{-2}$ and 500 mA $cm^{-2}$, respectively. These low overpotentials are close to the 34.3 mV at 10 mA $cm^{-2}$ of state-of-art Pt/C (20 %) but superior to 168.5 mV of Pt/C (20 %) at 100 mA $cm^{-2}$. Furthermore, the $V_3Si$ illustrates exceptional durability with no obvious decay in the 120 h at the different current densities (*i.e.*, 10 - 250 mA $cm^{-2}$). The excellent HER activity of $V_3Si$ alloy can be ascribed to the synergies of superior electronic conductivity and suitable *d*-band center. Moreover, DFT calculations reveal that $|\Delta G_{H*}|$ is decreased after introducing the V to Si. Beyond offering a stable and high-performance electrocatalyst in an acidic medium, this work inspires the rational design of desirable silicide electrocatalysts.

## 1. Introduction

Because of the growing demand for energy and concern about environmental pollution, the development of renewable sources has attracted considerable attention.[1-5] Among them, hydrogen is an ideal energy carrier owing to its high energy density and no pollution in hydrogen consumption.[6-8] Electrocatalytic water splitting is regarded as the most promising strategy to obtain high-purity hydrogen, which mainly relies on catalyst development.[9, 10] As a highly state-of-the-art electrocatalyst, platinum (Pt) has been reported to be promising to serve as a benchmark catalyst for HER. However, the natural scarcity and high cost limit its commercial applications.[11, 12] In this context, exploring affordable, durable, and highly efficient catalysts is still a significant challenge.

Recently, Si-group elements have gained increasing attention in electrocatalysis, considering the abundant Si-group elements on earth.[13-19] Constructing intermetallic compounds with ordered atomic arrangements and optimal electronic structures is a profitable approach to promote the electrocatalytic activity and durability of Si-group elements.[20-22] Significantly, introducing a second element with lower electronegativity to construct the donor-acceptor architectures can optimize the binding energy with intermediates, thus significantly improving the electrocatalytic performance.[23, 24] The transition metals (Au, Ti, Mn, Fe, Co, Ni, etc.) have been alloyed with Si and show excellent HER performance.[13, 19, 25, 26] Mu's group reported a series of intermetallic Pt-group silicides, including IrSi, PtSi, $Pd_2Si$, and RhSi, and possessed excellent HER catalysis activity.[15] Furthermore, Sun and his co-workers reported transition-metal-based intermetallic silicides (IrSi, RhSi, PtSi, $Ru_2Si_3$), intermetallic germanide ($Ru_2Ge_3$), and intermetallic stannides ($PdSn_3$, $Ru_3Sn_7$, $PdSn_2$, $IrSn_2$) prepared by a facile molten-salt-assisted strategy. The PtSi shows highly desirable electrocatalytic properties.[17] Although these pioneering works have been made, the Si-group intermetallic compounds for HER electrocatalysis have not been systematically studied.

Herein, we present an A15-type superconductor $V_3M$ (M = Si, Ge, and Sn) with high catalytic activity synthesizing by a facile strategy of arc melting. The activated $V_3Si$ superconductor catalyst displays a remarkable HER activity with a low overpotential of 33.4 mV at 10 mA $cm^{-2}$ and a Tafel slope of 32.2 mV $dec^{-1}$ in 0.5 M $H_2SO_4$ aqueous electrolyte. Moreover, it only requires 114.6 mV to reach a high current density of 500 mA $cm^{-2}$. Besides activity, the $V_3Si$ can operate at the current densities of 10 - 250 mA $cm^{-2}$ for 120 h without any noticeable decay. The synchrotron radiation spectroscopy and X-ray photoelectron spectroscopy (XPS) show significant electron transfer between the Si and V. The temperature-dependent electronic conductivities analysis and valence band spectra (VBS) illustrate the high electronic conductivity and suitable *d*-band center of $V_3Si$, respectively. The theoretical calculation reveals that introducing vanadium can decrease the $|\Delta G_{H*}|$, therefore facilitating hydrogen production. Remarkably, the $V_3Si$ superconductor only consists of abundant elements in the earth's crust compared with Pt and other Pt alternatives. These findings in our work may open noteworthy avenues to developing high-activation catalysts for V-based as well as rational design of Si-group electrocatalysts.

## 2. Experimental

### *2.1 Materials and reagents*

The vanadium powder (99.5% metals basis, ≥ 325 mesh) and germanium powder (99.99% metals basis) were purchased from Shanghai Aladdin Biochemical Technology Co., Ltd and Shanghai Macklin Biochemical Technology Co., Ltd respectively. The silica powder (crystalline, 99.5% metals basis, 100 mesh) and tin shot (3.175 mm, 99% metals basis) were obtained from Alfa Aesar. The isopropyl alcohol (AR) was purchased from Guangdong Guanghua Sci-Tech Co., Ltd., and sulfuric acid was purchased from Guangzhou Chemical Reagent Factory. In addition, Nafion binder (5 wt% in ethanol) was purchased from Sigma-Aldrich. All reagents were used as received. Carbon fiber paper (CeTech GDS250) was obtained from the SCI materials Hub.

### *2.2 Synthesis of $V_3M$ (M = Si, Ge, and Sn)*

In a typical process, the $V_3M$ (M = Si, Ge, and Sn) was prepared by using a metal melting furnace (MSM20-7) with a water-cooled copper hearth under a purified argon atmosphere according to the molar ratio. Before arc-melting, the V and Si or Ge were well-grounded in an agate mortar. Then, the powder-like mixtures were pressed into a thin cylinder with a radius of 6 mm. For the synthesis of $V_3Sn$, a tin ingot should be placed on top of vanadium that had been pressed into a thin cylinder due to the low melting point of tin. To improve the homogeneity of as-prepared samples, the samples were remelted three times by using a manipulator. The obtained $V_3Sn$ was further sealed inside an evacuated quartz tube and heated at 950 $^{o}$C for 6 h to get a pure phase. Finally, the target $V_3M$ (M = Si, Ge, and Sn) alloy balls were ground into powders for electrocatalyst.

*2.3 Material characterization*

The morphology was observed from the scanning electron microscope (SEM, COXEM EM-30AX), transmission electron microscopy (TEM, Hitachi HT7800), and high-resolution transmission electron microscopy (HRTEM, FEI Tecnai G2 F30, and JEOL JEM-2200FS). The X-ray diffraction (XRD) data were obtained by using Rigaku MiniFlex 600 (Cu Kα, $\lambda$ = 1.54178 Å). All XRD data are refined through the Rietveld method using FullProf Suite software to clarify the purity and structure of the samples. The surface chemical composition was acquired by X-ray photoelectron spectroscopy (XPS, Nexsa Thermo Fisher Scientific, USA). Electrical transport and i-V curves were recorded on a Quantum Design PPMS-14T by using a standard four-probe configuration.

*2.4 Measurements of Electrocatalytic HER*

To obtain the electrode for HER, the as-prepared $V_3Si$, $V_3Ge$, and $V_3Sn$ were sonicated in 1 mL isopropyl alcohol for 60 minutes. Then the mixture of supernatant (950 μL) and 5 % Nafion binder (50 μL) was sonicated for 10 minutes. Finally, the 50 μL of as-dispersed ink was dropped on the carbon fiber paper and dried naturally. The active area of carbon fiber paper is 0.25 cm$^2$. All electrochemical measurements were performed on the Multi Autolab M204 (Metrohm, Switzerland) at room temperature using a three-electrode system, in which the catalysts-coated carbon fiber paper, Pt sheet, and Ag/AgCl (Saturated KCl) electrode as working electrode, a counter electrode and reference electrode, respectively. All the polarization curves were carried out in an argon-saturated electrolyte of 0.5 M $H_2SO_4$ and converted to the reversible hydrogen electrode (RHE) by $E$ (RHE) = $E$ (Ag/AgCl) + 0.1989 V + 0.059 × $PH$ . The $iR$-correction was conducted according to the $E = E_{test} - i * 0.9 R_s$. The electrodes were first activated by cyclic voltammetry (CV) until stable electrochemical performances were achieved. The polarization curves were obtained by the linear sweep voltammetry (LSV) at the scan rate of 5 mV $s^{-1}$. The electrochemical impedance spectroscopy (EIS) was acquired at an amplitude of 5 mV in a frequency range from 100 kHz to 0.1 Hz. The electrochemical active surface area (ECSA) was characterized by the CV measurements in the potential range of non-Faradaic and the scan rates are 30, 50, 70, 90, and 100 mV $s^{-1}$. The ECSA value was calculated as follows.

$$ECSA = R_f \times S = \frac{C_{dl} \times S}{20\ \mu\text{F cm}^{-1}}$$

Where $R_f$ represents the roughness factor, $S$ is a geometric area of the electrode (0.25 $cm^2$), $C_{dl}$ represents the double value of the slop that plotted $\Delta j$ against the scan rate, and 20 μF $cm^{-2}$ is the $C_{dl}$ of the smooth metal surface.

In addition, the turnover frequency (TOF) was calculated based on the following:

$$\text{TOF} = \frac{j \times S}{2 \times F \times n}$$

Where $j$ (mA $cm^2$) is the current density, $S$ ($cm^2$) is the geometric area of the electrode, $F$ is Faraday’s constant (96485.3 C $mol^{-1}$), 2 is the number of transfer electrons, $n$ (mol) is the mole of the coated $V_3Si$ on the carbon fiber paper.

*2.5 Computational methods*

Density functional theory (DFT) calculations are performed within the Vienna Ab initio Simulation Package (VASP).[27, 28] The generalized gradient approximation (GGA) of the revised Perdew-Burke-Ernzerhof (RPBE) method is applied to calculate the adsorption energies between the adsorbate and catalyst surface.[29, 30] A vacuum space of 15 Å is adopted in the surface slab to avoid interactions between adjacent images. An energy convergence criterion of $10^{-5}$ eV between two ionic steps is adopted. The energy cut-off is set to 400 eV, and Γ-centered k-point meshes of 4 ×2 ×1 and 2 ×2 ×1 are used for the (2 ×2) supercell slab of $V_3Si$ and Si, respectively.

## 3. Results and Discussion

As a typical early transition metal with multiple valence states, vanadium (V) possesses an adjustable electronic structure that is mainly responsible for optimizing the catalytic activity of the host material.[31] Moreover, V's low cost and abundant resources make it promising for commercialization.[32] Therefore, the V is introduced to the alloy with Si-group elements to investigate the HER electrocatalysis of Si-group intermetallic compounds systematically. We then synthesize the $V_3M$ (*i.e.*, $V_3Si$, $V_3Ge$, and $V_3Sn$) superconductors via a facile technology of arc melting. The synthesis processes are shown in Figure 1a, and stoichiometric mixtures of V and Si-group elements (*i.e.*, Si, Ge, and Sn) form an alloy by arc-melting in the Ar atmosphere.

Powder X-ray diffraction is first performed to investigate the crystalline phase of the as-prepared Si-group samples. Figures 1b-1d depict the X-ray diffraction (XRD) patterns and the corresponding refinement data of $V_3M$ (M = Si, Ge, and Sn). The obvious diffractions correspond to $V_3Si$ (ICDD/PDF No. 04-005-8937),[33] $V_3Ge$ (ICDD/PDF No. 01-072-1967),[34] and $V_3Sn$ (ICDD/PDF No. 04-002-9958).[35] The prominent diffraction peaks around 38.0°, 42.7°, and 47.0° can be indexed to (200), (210), and (211) facets of the $V_3M$ (*i.e.*, $V_3Si$, $V_3Ge$, and $V_3Sn$) samples. A Rietveld refinement of the XRD patterns reveals that the prepared intermetallic $V_3M$ (*i.e.*, $V_3Si$, $V_3Ge$, and $V_3Sn$) is of good quality. Detailed refinement results and crystal structures are summarized in Tables S1 and S2. It shows the $V_3M$ (*i.e.*, $V_3Si$, $V_3Ge$, and $V_3Sn$) possesses the A15 structure with a space group of *Pm*-3*n*, where the Si, Ge, or Sn atoms form a body-centered-cubic unit cell, and each surface of the cubic contains two V atoms (shown in the inset of Figure 1b, 1c and 1d). Moreover, the lattice constant *a* is increased with the increase of atomic number, namely, 4.73 Å, 4.78 Å, and 4.99 Å for $V_3Si$, $V_3Ge$, and $V_3Sn$, respectively. The different distances may result in the various charge transfer abilities between the V and the Si, Ge, or Sn.[36]

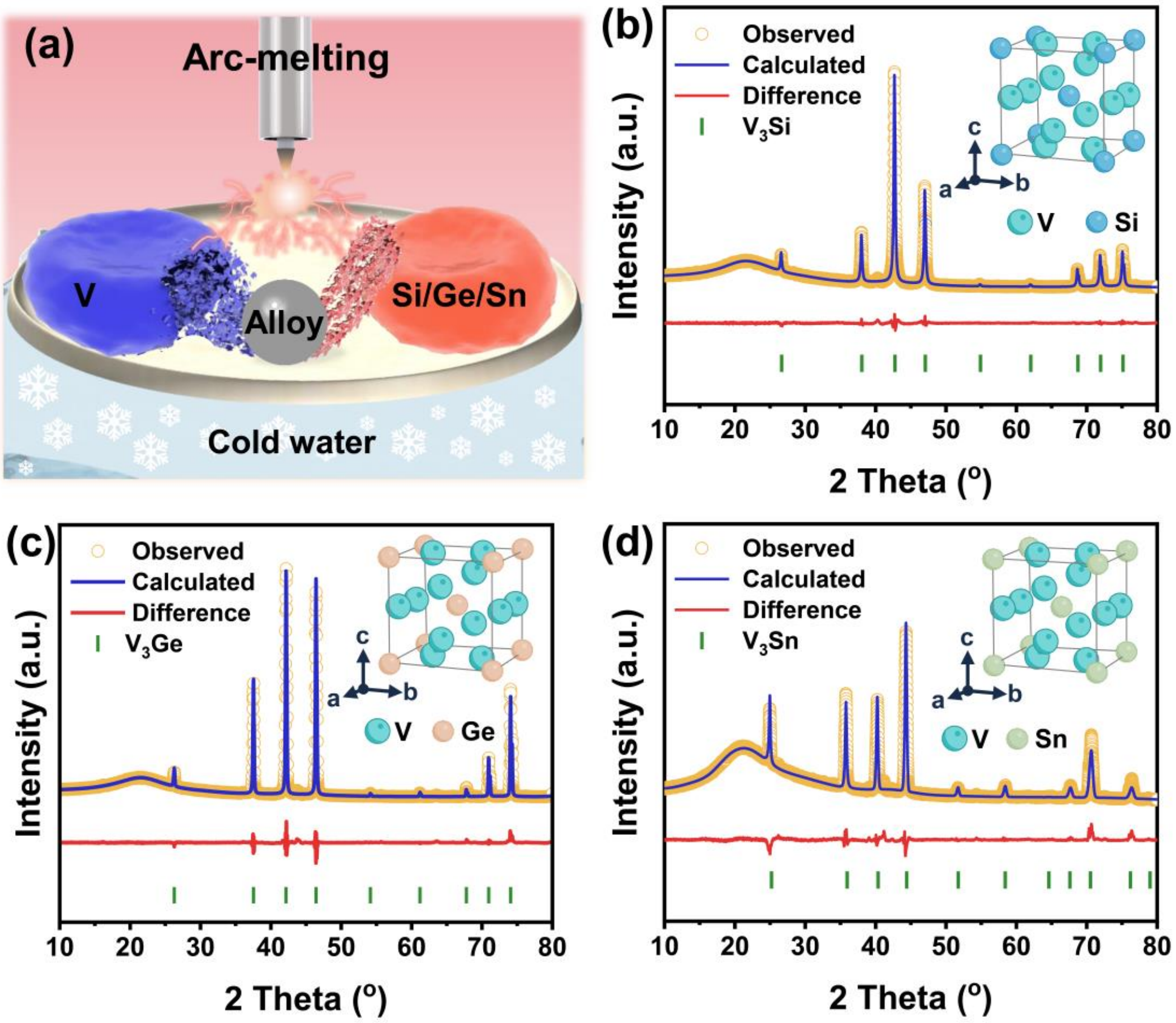

**Figure 1**. Preparation and phase analysis of $V_3M$ (*i.e.*, $V_3Si$, $V_3Ge$, and $V_3Sn$) superconductor samples. (a) Schematic illustration of the fabrication of $V_3M$ (*i.e.*, $V_3Si$, $V_3Ge$, and $V_3Sn$) through the facile method of arc-melting. Observed and calculated XRD patterns of as-prepared (b) $V_3Si$, (c) $V_3Ge$, and (d) $V_3Sn$. The inserts are the crystal structure of the $V_3M$ (*i.e.*, $V_3Si$, $V_3Ge$, and $V_3Sn$) with a space group of *Pm*-3*n*.

As a proof-of-concept application, the HER performances of $V_3M$ (*i.e.*, $V_3Si$, $V_3Ge$, and $V_3Sn$) superconductor compounds are investigated in a standard three-electrode system with 0.5 M $H_2SO_4$, and the results are presented in Figure 2. The linear sweep voltammetry (LSV) polarization curves (Figure 2a) after activation processes (Figure S1) illustrate that the $V_3Si$ shows superior electrocatalytic activity than the $V_3Ge$ and $V_3Sn$. To drive a current density of 10 mA $cm^{-2}$, $V_3Si$ requires an overpotential of 33.4 mV, which is much lower than that of $V_3Ge$ (*i.e.*, 148.3 mV) and $V_3Sn$ (*i.e.*, 182.3 mV). The Tafel slopes are calculated based on the LSV curves according to the equation of $\eta = b \log j + a$, where $b$ is the Tafel slope to further judge the reaction mechanism of HER activity. The lowest value is calculated to be 32.2 mV $dec^{-1}$ for $V_3Si$, while the highest Tafel slope is found for $V_3Sn$ (*i.e.*, 126.6 mV $dec^{-1}$). These results indicate that the HER occurs through a Volmer-Heyrovsky mechanism, and the rate-limiting step is the electrochemical recombination with an extra proton (Figure 2b)[37]. At an overpotential of 100 mA $cm^{-2}$, comparing the overpotential and Tafel slope shows that the $V_3Si$ electrode outperforms the $V_3Ge$ and $V_3Sn$ electrodes (Figure 2c). The electrochemical impedance spectra (EIS) show that the $V_3Si$ processes the smallest charge-transfer resistance ($R_{ct}$) compared with the $V_3Ge$ and $V_3Sn$, suggesting that the $V_3Si$ has faster charge-transfer kinetics (Figure S2). The electrocatalyst electrochemically active surface area (ECSA) is calculated based on the CV measurements (Figure S3-S5). The largest ECSA value of $V_3Si$ means that it has more active sites than $V_3Ge$ and $V_3Sn$. To further evaluate the intrinsic activity of $V_3M$ (*i.e.*, $V_3Si$, $V_3Ge$, and $V_3Sn$), the specific activities derive from the current density normalized by ECSA. As depicted in Figure 2d, at $\eta = 0.1$ V, the $V_3Si$ possesses a high current density of 112.69 mA $cm^{-2}{}_{ECSA}$, up to 55.0 and 60.6 times higher than $V_3Ge$ and $V_3Sn$, respectively. The turnover frequencies (TOFs) of $V_3M$ (*i.e.*, $V_3Si$, $V_3Ge$, and $V_3Sn$) are further calculated to reveal their intrinsic activities. Remarkably, at an overpotential of 100 mV, $V_3Si$ shows the largest TOF value (1.34 $H_2$ $s^{-1}$), 110.7 and 111.8 times than $V_3Ge$ and $V_3Sn$ respectively (Figure 2e). These results evidence that the $V_3Si$ has considerably enhanced intrinsic HER activity.

The XRD, inductively coupled plasma emission spectrometer (ICP), and TEM measurements are carried out to understand the influence of the activation process on the structure. As depicted in Figure S6, the current density of the initial state of $V_3Si$ is much worse, while it illustrates a stable HER performance after 5000 CV cycles. The XRD test shows no other detectable peaks, illustrating the crystal structure is maintained after the activation process of $V_3Si$ (Figure S7). We suppose that it can be attributed to the exfoliation of bulk $V_3Si$ and exposure to more activated sites, thus improving the catalytic performance. To verify this hypothesis, the ICP and TEM measurements are conducted to measure the electrode material and electrolyte after 6000 CV cycles, respectively. As shown in Figure S8, the presence of the V and Si elements in the electrolyte can be confirmed. Furthermore, the TEM images show a thinner morphology after cycling, as shown in Figure S9. These results suggest the presence of exfoliation of bulk $V_3Si$ during the cycling.

The high HER activity of the $V_3Si$ superconductor is further evaluated by comparing it with those of Pt/C (20%). As shown in Figure 2a, to reach the current density of 10 mA cm$^{-2}$, the $V_3Si$ superconductor delivers a small overpotential requirement of 33.4 mV, slightly smaller than that of Pt/C (20%) (*i.e.*, 34.3 mV). Nevertheless, at the higher current density, the overpotential of the $V_3Si$ superconductor is much lower than that of Pt/C (20%), suggesting that the high intrinsic activity of $V_3Si$ superconductor ensures fast mass and electron transport at large current density (Figure 2f). With high catalytic activity, the $V_3Si$ superconductor only needs 114.6 mV to realize a high current density of 500 mA cm$^{-2}$. As shown in Figure 2b, the Tafel slope of 32.2 mV dec$^{-1}$ is superior to the commercial Pt/C (20%). Apart from the activity, the stability of electrocatalysts is vital for the practical application. The long-term chronoamperometry curves reveal that the $V_3Si$ superconductor can be stability operated between the current density of 10 - 250 mA cm$^{-2}$ without obvious decay (Figure 2g). In contrast, the Pt/C (20%), $V_3Ge$, and $V_3Sn$ exhibit a rapid current decay in the initial few hours. Moreover, the excellent HER activity of the $V_3Si$ superconductor is superior to that of most reported Si-based HER electrocatalysts (Figure 2h and Table S3) and precedes most of the reported acidic HER catalysts (Table S4).[13, 15, 19, 25, 31, 38-43]

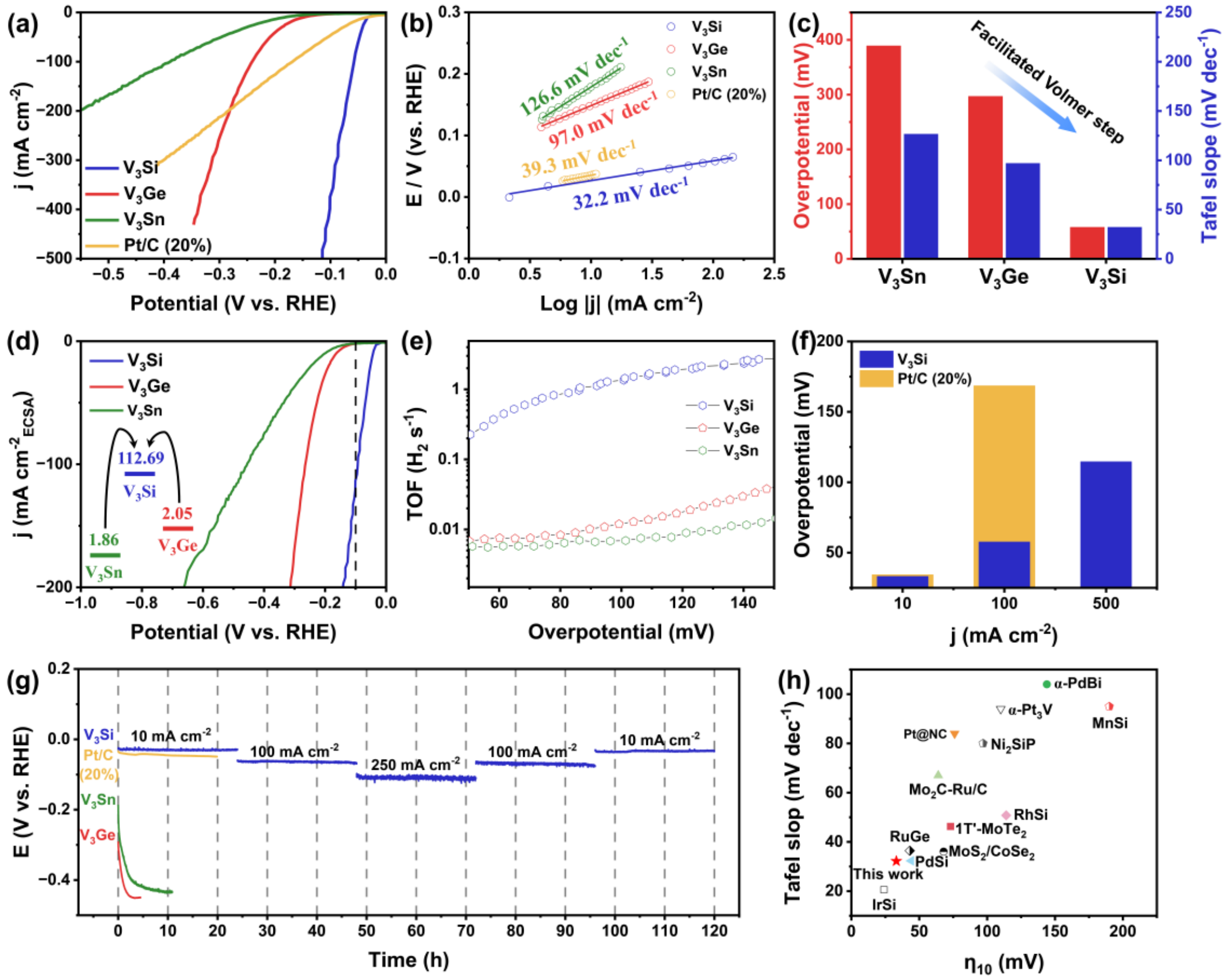

**Figure 2**. HER electrocatalytic properties of the $V_3M$ (*i.e.*, $V_3Si$, $V_3Ge$, and $V_3Sn$) catalysts. (a) The *iR*-corrected LSV curves and (b) corresponding Tafel slopes. (c) The comparison of Tafel slopes and catalytic activity in terms of the overpotential at 100 mA cm$^{-2}$. (d) Specific activities via normalizing the HER current by ECSA. (e) TOF values of the $V_3M$ (*i.e.*, $V_3Si$, $V_3Ge$, and $V_3Sn$) superconductors. (f) Comparison of overpotential of $V_3Si$ and Pt/C (20%) between 10 - 500 mA cm$^{-2}$. (g) The cycling stability of $V_3Si$, $V_3Ge$, $V_3Sn$, and Pt/C (20%) catalysts. (h) Comparison of the HER activities for $V_3Si$ superconductor and the reported electrocatalysts.

To obtain more profound insights into the origin of outstanding HER catalytic activity of $V_3Si$, the structural/component characterizations are performed. First, the temperature-dependent resistivity measurement is performed to reveal the superconductivity of $V_3M$ (*i.e.*, $V_3Si$, $V_3Ge$, and $V_3Sn$). It can be seen that the normalized resistivity ($\rho/\rho$300K) vs temperature, as shown in Figure 3a, insulates that the superconducting critical transition temperature ($T_c$) $V_3Si$ is as high as 17.0 K and the $T_c$ of $V_3Ge$ and $V_3Sn$ are 4.2 K and 7.0 K respectively, which are consistent with the previous report [44-46]. As superconductors, the resistivities of $V_3M$ (*i.e.*, $V_3Si$, $V_3Ge$, and $V_3Sn$) decrease with the decreasing temperature, which means it may operate at low temperatures. After grinding the alloy ball, microparticles of the $V_3Si$ sample can be obtained in Figure S10. Besides, elemental mapping analysis displays that V and Si elements are evenly distributed, and the element ratio between them is close to 3: 1. To reveal the feature of microscopic crystallography, the high-resolution transmission electron microscope (HRTEM) is applied subsequently. From the HRTEM image, as shown in Figures 3b and 3c, the $V_3Si$ superconductor possesses multiple layers, and the observed lattice fringes with *d*-spacing of 0.198 and 0.218 nm can be described to (211) and (210) plane of $V_3Si$. After further magnifying, it can be seen that the cubic $V_3Si$ displays the order arrangement of V and Si atoms from the STEM image Figure 3d. Moreover, the energy-dispersive X-ray spectroscopy (EDS) mappings confirm that the V and Si elements are evenly throughout $V_3Si$ without apparent separation or aggregation (Figure 3e).

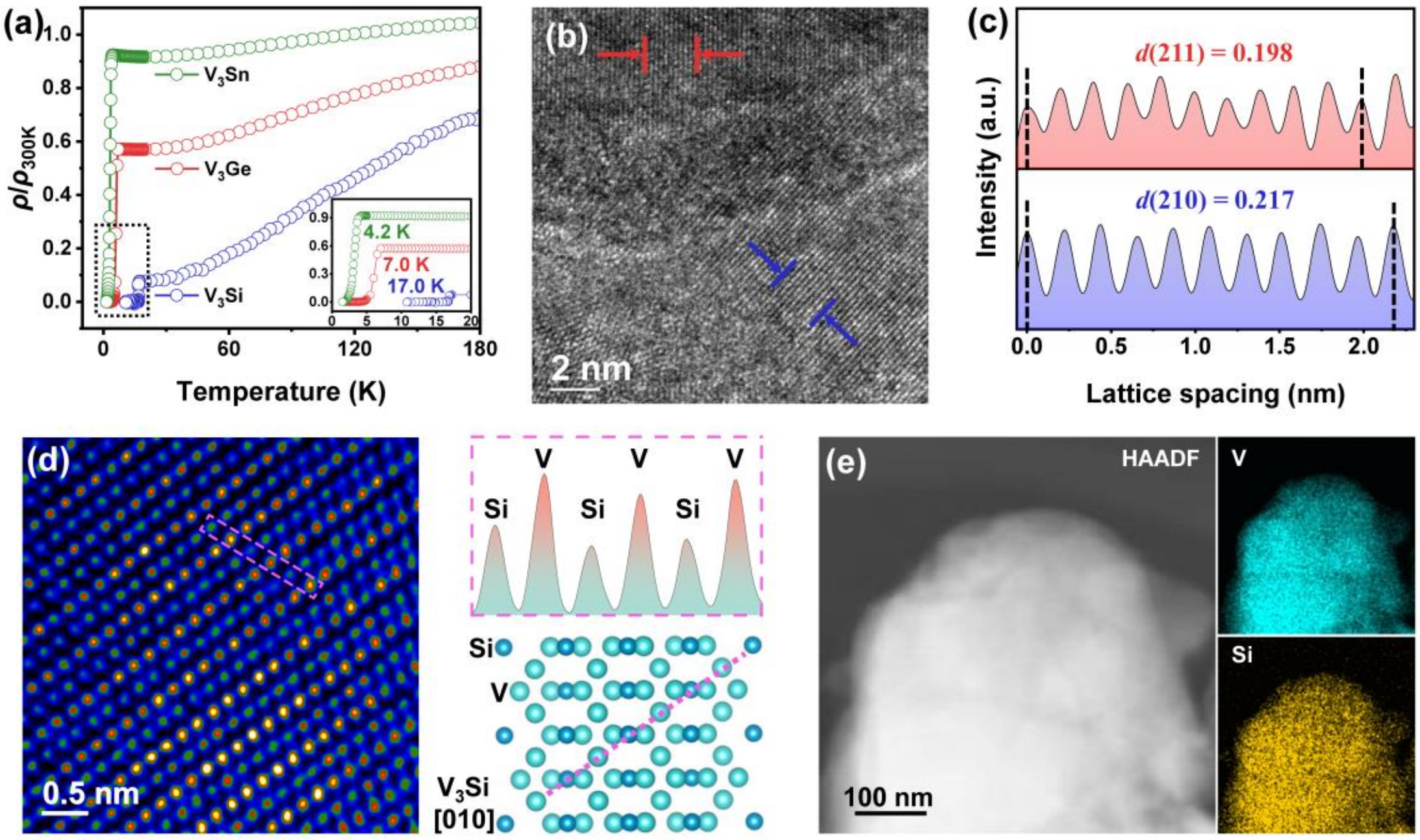

**Figure 3**. (a) The overview of the temperature dependence of normalized resistivity (ρ/ρ300 K) for the $V_3M$ (*i.e.*, $V_3Si$, $V_3Ge$, and $V_3Sn$) electrocatalysts. Inset is the enlarged portion (0 to 20 K). (b) HRTEM image and (c) lattice fringes with *d*-spacing of $V_3Si$ superconductor. (d) HAADF−STEM image for $V_3Si$ superconductor. (e) Element mappings of V and Si for $V_3Si$ superconductor.

To further understand the chemical environment and electronic states of V and Si elements, X-ray photoelectron spectroscopy (XPS), X-ray absorption near-edge structure (XANES), and extended X-ray absorption fine structure (EXAFS) are carried out. As shown in Figures 4a and 4b, the binding energy of Si 2p for $V_3Si$ displays a positive shift compared with the Si, while the negative shift in the binding energy of V 2p can be observed when $V_3Si$ is compared with V. This indicates a significant change in the electronic structure of Si after introducing the V to form the $V_3Si$.[15] Notably, XPS verifies that the V and Si in $V_3Si$ exist mainly in the metallic state despite an oxidation state due to exposure to air. The stronger intensity of the oxidation signal of Si in $V_3Si$ derived from the surface oxidized silicon species (Figure S11).[14, 17] As depicted in Figure S12, the XPS results of V3Ge and V3Sn reveal the metallic state's existence.[47, 48] These results indicate $V_3M$ (*i.e.*, $V_3Si$, $V_3Ge$, and $V_3Sn$) is an alloy compound and chemical states of V, Ge, Sn, and Si exbibit metallic state. Moreover, as illustrated in XANES spectra, the absorption edge position of V in the $V_3Si$ superconductor is both lower than those of $V_2O_3$, $V_2O_5$, and V foil, indicating the electron-enrich state of V atoms in the $V_3Si$ superconductor (Figure 4c). For the pre-edge, a much lower intensity of $V_3Si$ superconductor than that of $V_2O_3$ and $V_2O_5$ implies lower coordination due to the pre-edge intensities and coordination deviation of the centrosymmetric photo absorbers being in direct ratio. The normalized EXAFS spectroscopy is carried out further to investigate the local environment and coordination structure of V. It can be seen that the identified peaks for $V_3Si$ are located at 2.37 Å and 4.85 Å belonging to the V-Si bond. The V-Si bond is significantly different from the V-V bond (~2.27 Å) in V foil and the V-O bond (~1.53 Å) in $V_2O_3$ and $V_2O_5$ (Figure 4d). Moreover, wavelet transformed (WT)-EXAFS analysis is applied to present this information intuitively (Figure 4e). The wavelet data near 13.0 $Å^{-1}$ can be ascribed to the V-Si bond for $V_3Si$, which is prominently different from the 7.4 $Å^{-1}$ of the V-V bond and 5.8 $Å^{-1}$ of the V-O bond. These results indicate that two coordination environments of V-V and V-Si coordination exist in the $V_3Si$ superconductor. More importantly, these results of XPS and XANES mean that existing change transfers from Si to V.

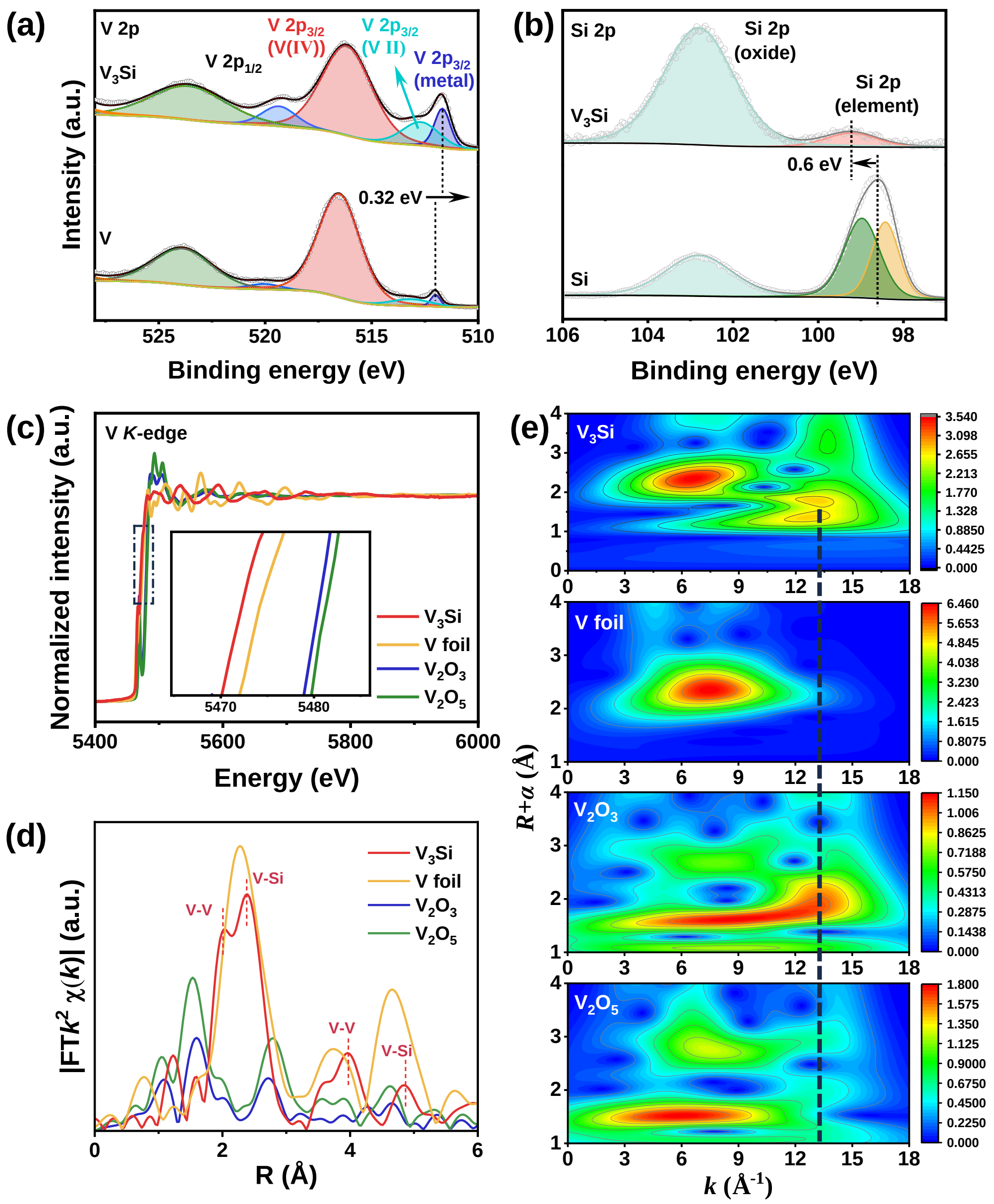

**Figure 4**. Atomic structure characterizations of $V_3Si$ superconductor. (a) V 2p and (b) Si 2p high-resolution XPS spectra for $V_3Si$. (c) Experimental V K-edge XANES spectra of $V_3Si$ superconductor, $V_2O_5$, $V_2O_3$, and V foil, respectively. (d) Fourier transforms of the EXAFS spectra of $V_3Si$ superconductor and reference samples. (e) WT-EXAFS plots of $V_3Si$ superconductor and along with reference samples.

The direct-current electric transport test, valence band spectra (VBS), and first-principles density functional theory (DFT) calculations are conducted to probe the reaction energetics further. As depicted in Figure 5a and Figure S13, the electronic conductivity measurements reveal that the conductivity of the $V_3M$ (*i.e.*, $V_3Si$, $V_3Ge$, and $V_3Sn$) increases with a decrease in temperature. Especially in the range of room temperature (the pink region, 15 - 30 $^oC$), the $V_3Si$ delivers the highest electronic conductivity among $V_3Si$, $V_3Ge$, and $V_3Sn$ (Figure 5b). The higher electronic conductivity is beneficial for the charge transfer between reactant molecules and electrodes and thus improves the electrocatalysis performance. As one of the most successful theories, the *d*-band center is closely related to the catalytic activity of catalysts. As illustrated in valence band spectra (VBS) of Figure 5c, the *d*-band center of $V_3Si$, $V_3Ge$, and $V_3Sn$ are calculated to be 4.34 eV, 5.68 eV, and 5.48 eV, respectively. According to the *d*-band center theory, the higher *d*-band center means antibonding orbits are occupied by more electrons, further weakening the binding mode between the catalyst surface and the $H^*$.[49] Therefore, the suitable *d*-band center of $V_3Si$ is responsible for its enhanced HER activity. Subsequently, DFT calculations are performed to better understand the excellent HER performance of V3Si. After considering all possible adsorption configurations, one can get the energetically most favorable sites of the $V_3Si$-(210) surface and the Si-(111) surface, which are shown in Figure S14. Based on this, the hydrogen adsorption Gibbs free energy ($\Delta G_{H*}$), a well-accepted descriptor for HER, is calculated. The Si-(111) shows a remarkably low Gibbs free energy, indicating a strong binding energy with the hydrogen atoms. After the introduction of vanadium, the adsorption strength of $V_3Si$ is decreased compared with that of Si, leading to a $\Delta G_{H*}$ closer to zero, as shown in Figure 5d. In other words, introducing V to Si would favor the transformation from $H^*$ to $H_2$ and expedite the $H_2$ desorption to refresh the catalytic active sites. In summary, the superior HER activity of $V_3Si$ alloy can be described to the following reasons: a) The high electronic conductivity of $V_3Si$ ensures the fast charge transfer between $H^*$ and the catalyst surface. d) The suitable *d*-band center weakens the binding mode between $H^*$ and the catalyst surface and thus enhances the catalysis performance.

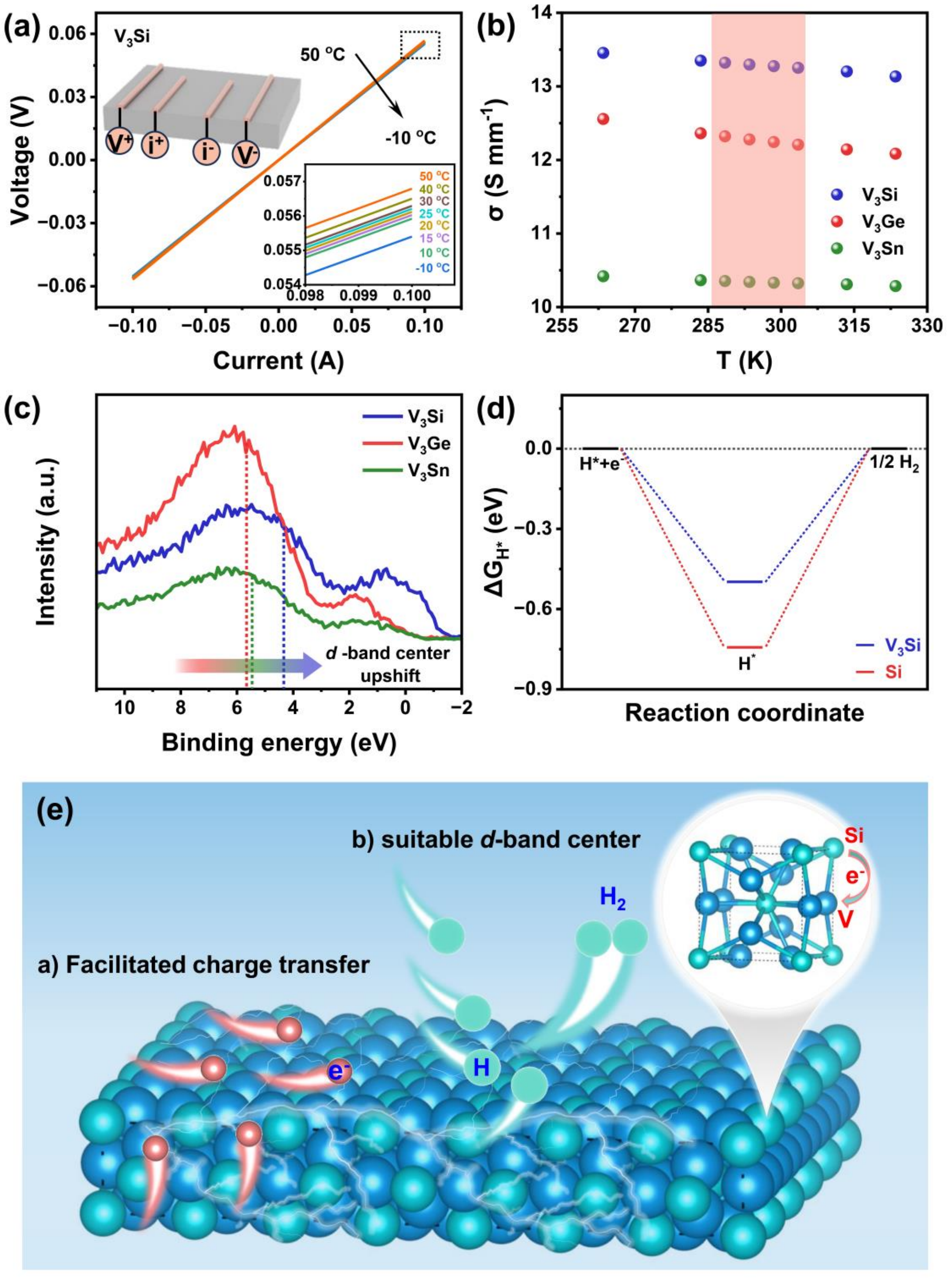

**Figure 5.** Investigation of electrocatalytic mechanism. (a) I-V curves as a function of the temperature of $V_3Si$. The insert is the illustration of electrical transport tests. (b) The temperature-dependent electronic conductivities analysis for $V_3M$ (*i.e.*, $V_3Si$, $V_3Ge$, and $V_3Sn$). (c) Valence band spectra (VBS) of $V_3M$ (*i.e.*, $V_3Si$, $V_3Ge$, and $V_3Sn$) alloys as measured by high-resolution X-ray photoelectron spectroscopy. (d) Reaction coordinates of $V_3Si$ and Si in the process of HER. (e) Diagram showing HER catalytic reaction in $V_3Si$.

## 4. Conclusion

To summarize, we successfully synthesized a series of A15-type superconductors by arc-melting. Compared with $V_3Ge$ and $V_3Sn$, the $V_3Si$ only needs low overpotentials of 33.4 mV and 114.6 mV to reach 10 mA $cm^{-2}$ and 500 mA $cm^{-2,}$ respectively, which is much lower than these of $V_3Ge$ (*i.e.*, 148.5 at 10 mA $cm^{-2}$ and 241.1 mV at 100 mA $cm^{-2}$) and $V_3Sn$ (182.4 mV at 10 mA $cm^{-2}$ and 388.7 at 100 mA $cm^{-2}$). Moreover, the $V_3Si$ exhibits superior cycling stability at the current density of 10 - 250 mA $cm^{-2}$ for 120 h. The excellent catalysis performance of $V_3Si$ alloy profits from the optimization of electronic structures of $V_3Si$, resulting in higher electronic conductivity and a suitable *d*-band center. The discovery of this $V_3Si$ superconductor catalyst paves the way for the promising application of quantum material in energy conversion and highlights the rational design of desirable silicide electrocatalysts.

**CRediT authorship contribution statement**

**Peifeng Yu**: Conceptualization, Methodology, Investigation, Validation, Formal analysis, Writing - original draft; **Jie Zhan**: Conceptualization, Methodology, Investigation, Formal analysis, Writing - original draft; **Xiaobin Zhang:** TEM characterization; **Kangwang Wang**: Discussion, Writing - review & editing; **Lingyong Zeng**: Discussion, Writing - review & editing; **Kuan Li**: Discussion, Writing - review & editing; **Chao Zhang**: Discussion, Writing - review & editing; **Longfu Li**: Discussion, Writing - review & editing; **Ying Liang**: Writing - review & editing; **Kai**

**Yan**: Conceptualization, Writing - review & editing; **Yan Sun**: Conceptualization, Writing - review & editing; **Huixia Luo**: Resources, Conceptualization, Writing - review & editing, Supervision, Project administration, Funding acquisition. Peifeng Yu and Jie Zhan contributed equally to this work.

## Declaration of Competing Interest

The authors declare no competing interests.

## Acknowledgments

This work was supported by the National Natural Science Foundation of China (12274471, 11922415), Guangdong Basic and Applied Basic Research Foundation (2022A1515011168) and Key Research & Development Program of Guangdong Province, China (2019B110209003). Yan Sun acknowledges supports from the National Natural Science Foundation of China (Grant No.52271016), Liaoning Province (Grant No. XLYC2203080). Part of the numerical calculations in this study were carried out on the ORISE Supercomputer (Grants No. DFZX202319). The experiments reported were conducted at the Guangdong Provincial Key Laboratory of Magnetoelectric Physics and Devices, No. 2022B1212010008.

## Data availability

Data will be made available on request.

# Supporting Information

## Mapping Hydrogen Evolution Activity Trends of V-based A15 Superconducting Alloys

Peifeng Yu [a, 1], Jie Zhan [e, f, 1], Xiaobing Zhang [b], Kangwang Wang [a], Lingyong Zeng [a], Kuan Li [a], Chao Zhang [a], Longfu Li [a], Ying Liang [c], Kai Yan [d *], Yan Sun [e, f *], Huixia Luo [a *]

[a] School of Materials Science and Engineering, State Key Laboratory of Optoelectronic Materials and Technologies, Guangdong Provincial Key Laboratory of Magnetoelectric Physics and Devices, Key Lab of Polymer Composite & Functional Materials, Sun Yat-Sen University, Guangzhou 510275, China

[b] State Key Laboratory for Advanced Metals and Materials, University of Science and Technology Beijing, Beijing 100083, China

[c] Basic course department, Guangzhou Maritime University, Guangzhou 510725, China

[d] School of Environmental Science and Engineering, Sun Yat-Sen University, Guangzhou 510275, China

[e] Shenyang National Laboratory for Materials Science, Institute of Metal Research, Chinese Academy of Sciences, Shenyang, 110016, China

[f] School of Materials Science and Engineering, University of Science and Technology of China, Shenyang 110016, China

[1] These authors contributed equally

*Corresponding author:

Prof. Kai Yan (yank9@mail.sysu.edu.cn)

Prof. Yan Sun (sunyan@imr.ac.cn)

Prof. Huixia Luo (luohx7@mail.sysu.edu.cn)

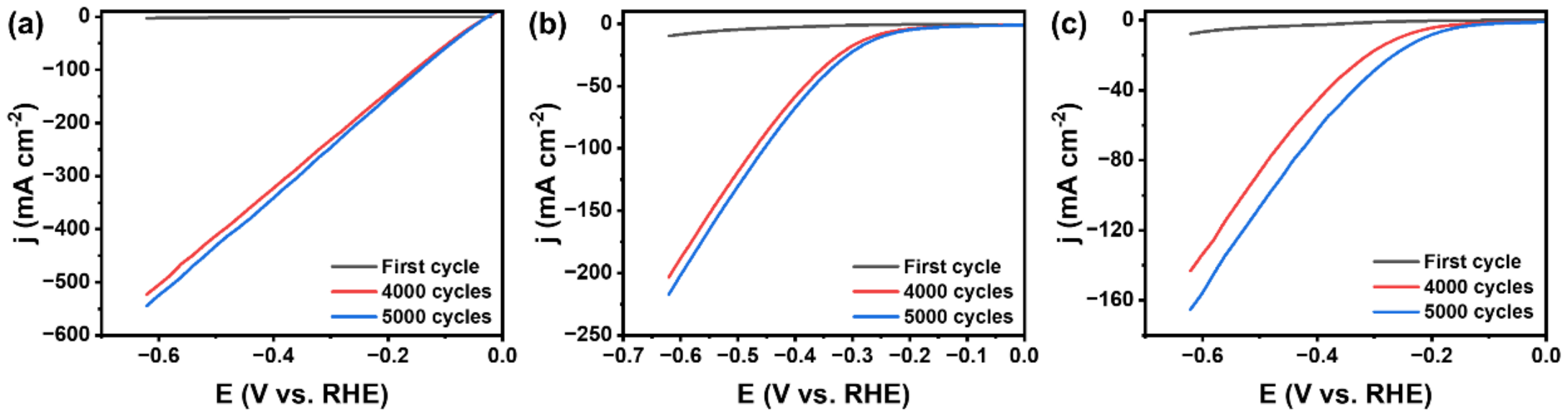

**Figure S1**. LSV curves without *iR*-correstion for (a) $V_3Si$, (b) $V_3Ge$, and (c) $V_3Sn$ at different cycles at a scan rate of 100 mV $s^{-1}$.

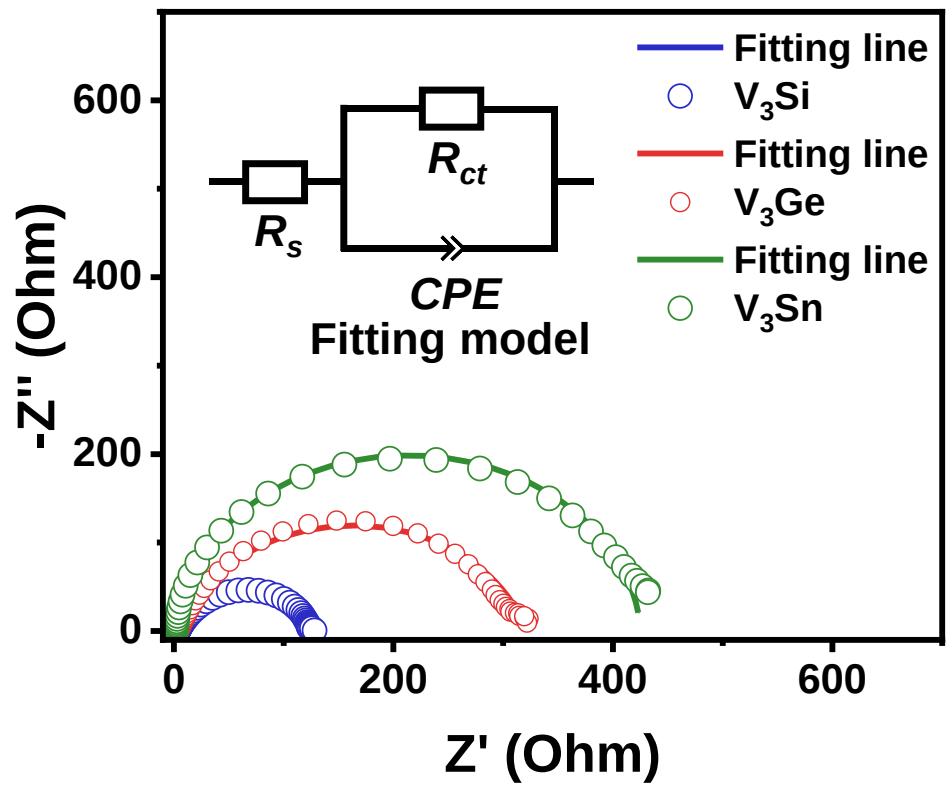

**Figure S2.** EIS of $V_3Si$, $V_3Ge$ and $V_3Sn$. The inset shows the corresponding equivalent circuit model.

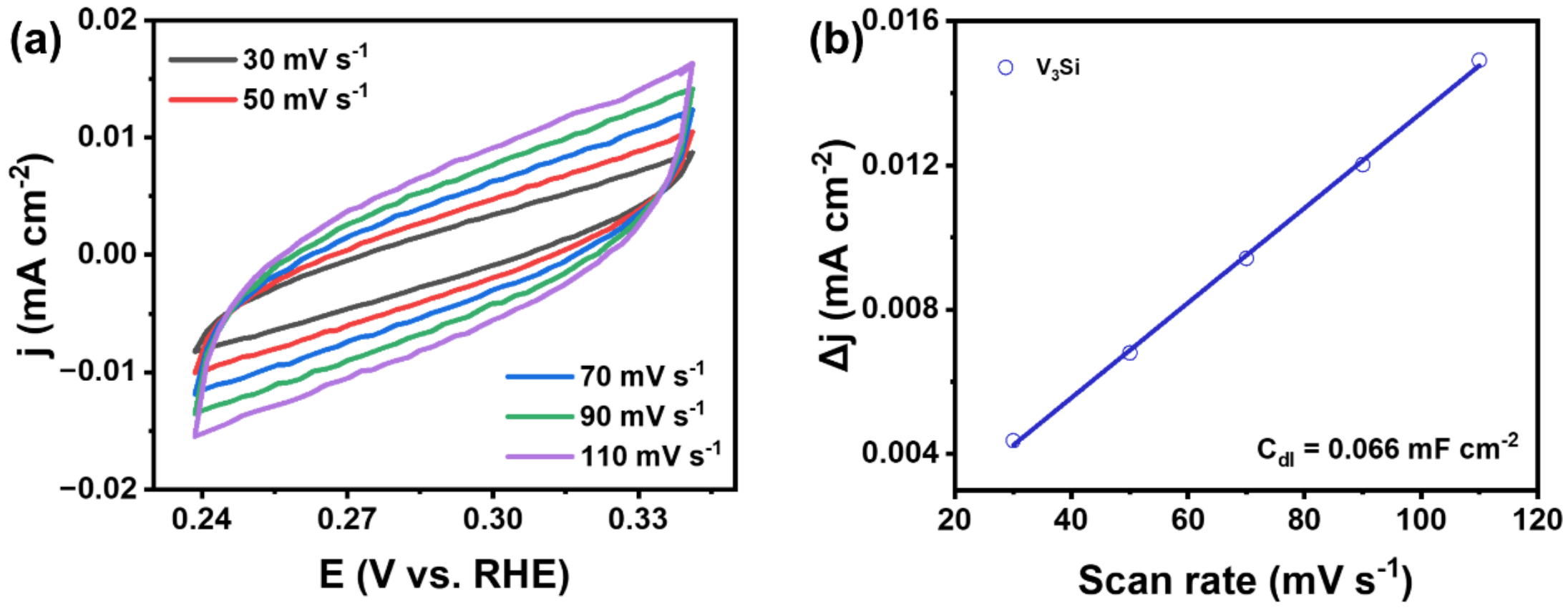

**Figure S3**. CV curves of $V_3Si$ electrode measured at 30, 50, 70, 90, and 110 mV $s^{-1}$.

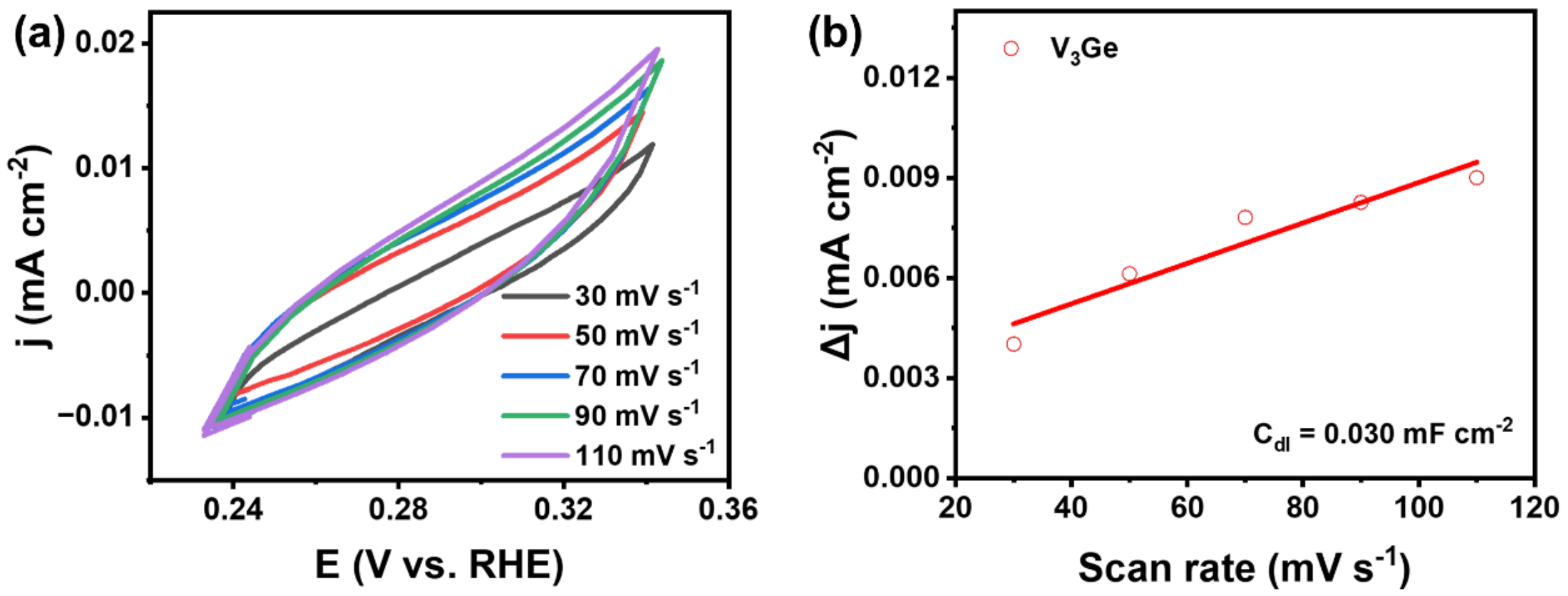

**Figure S4**. CV curves of $V_3Ge$ electrode measured at 30, 50, 70, 90, and 110 mV $s^{-1}$.

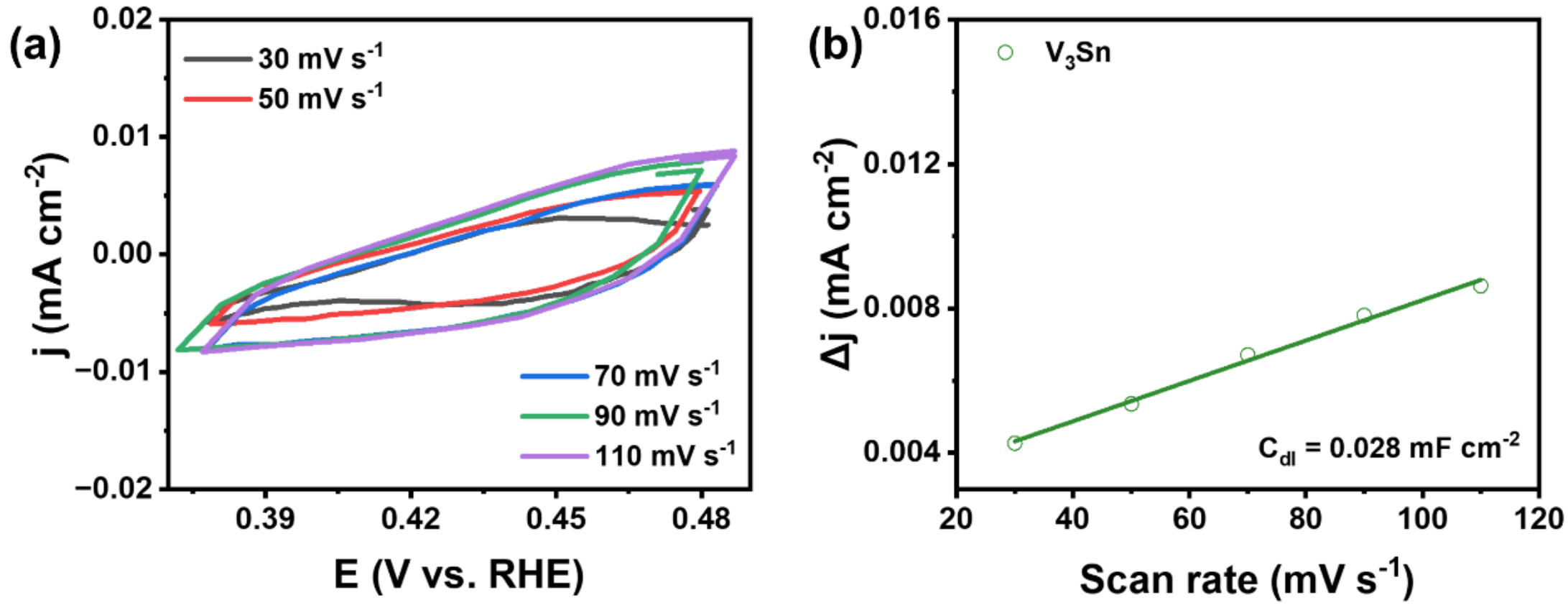

**Figure S5**. CV curves of $V_3Sn$ electrode measured at 30, 50, 70, 90, and 110 mV $s^{-1}$.

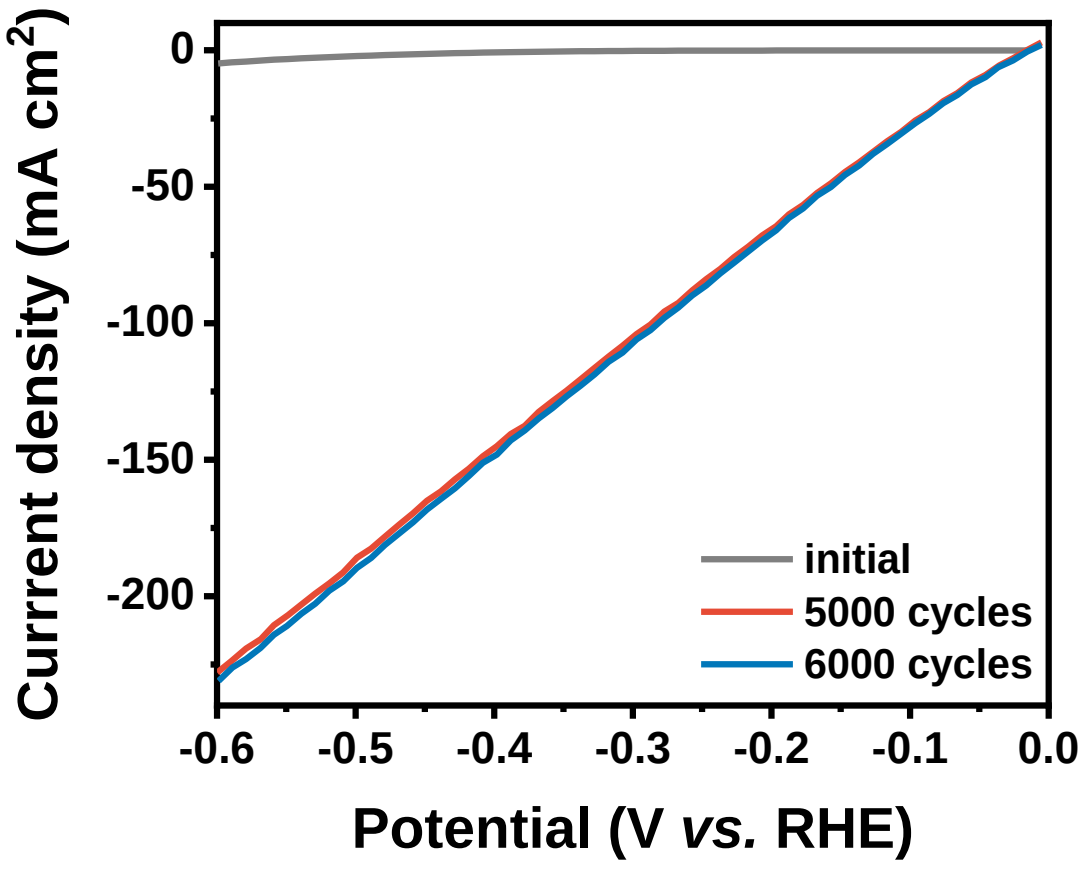

**Figure S6**. CV measurements for $V_3Si$ superconductor at different cycles at a scan rate of 100 mV $s^{-1}$.

For detecting the structure of $V_3Si$ superconductor coated on the carbon fiber paper, we used the carbon fiber paper with 1 * 1 cm as the substrate and dropped 950 μL of supernatant onto the carbon fiber paper.

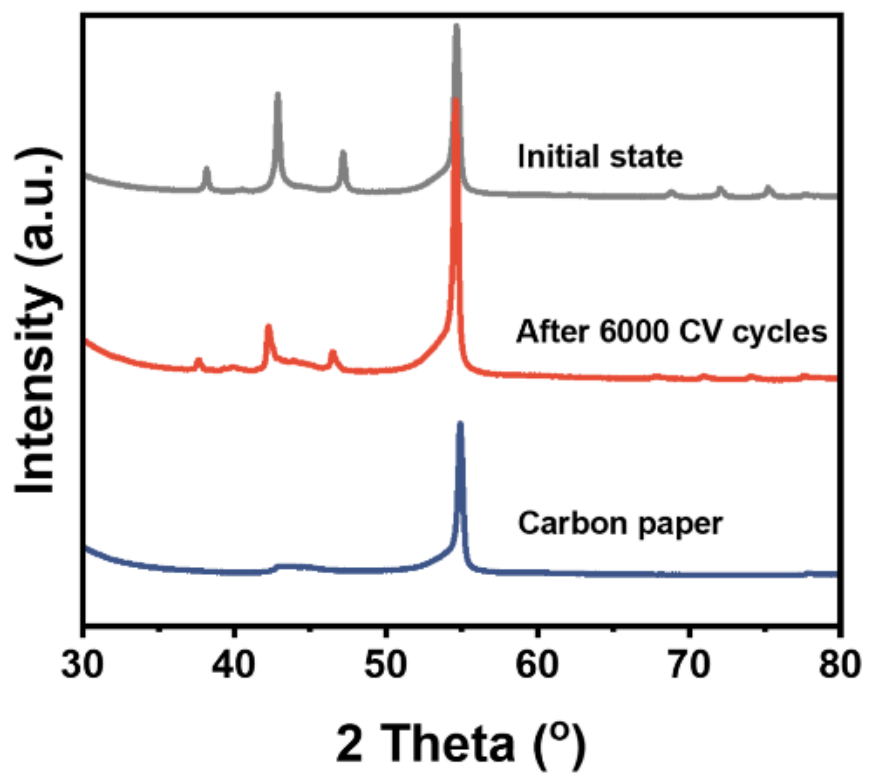

**Figure S7**. XRD patterns for carbon paper, the electrode of the initial, the electrode after activation process.

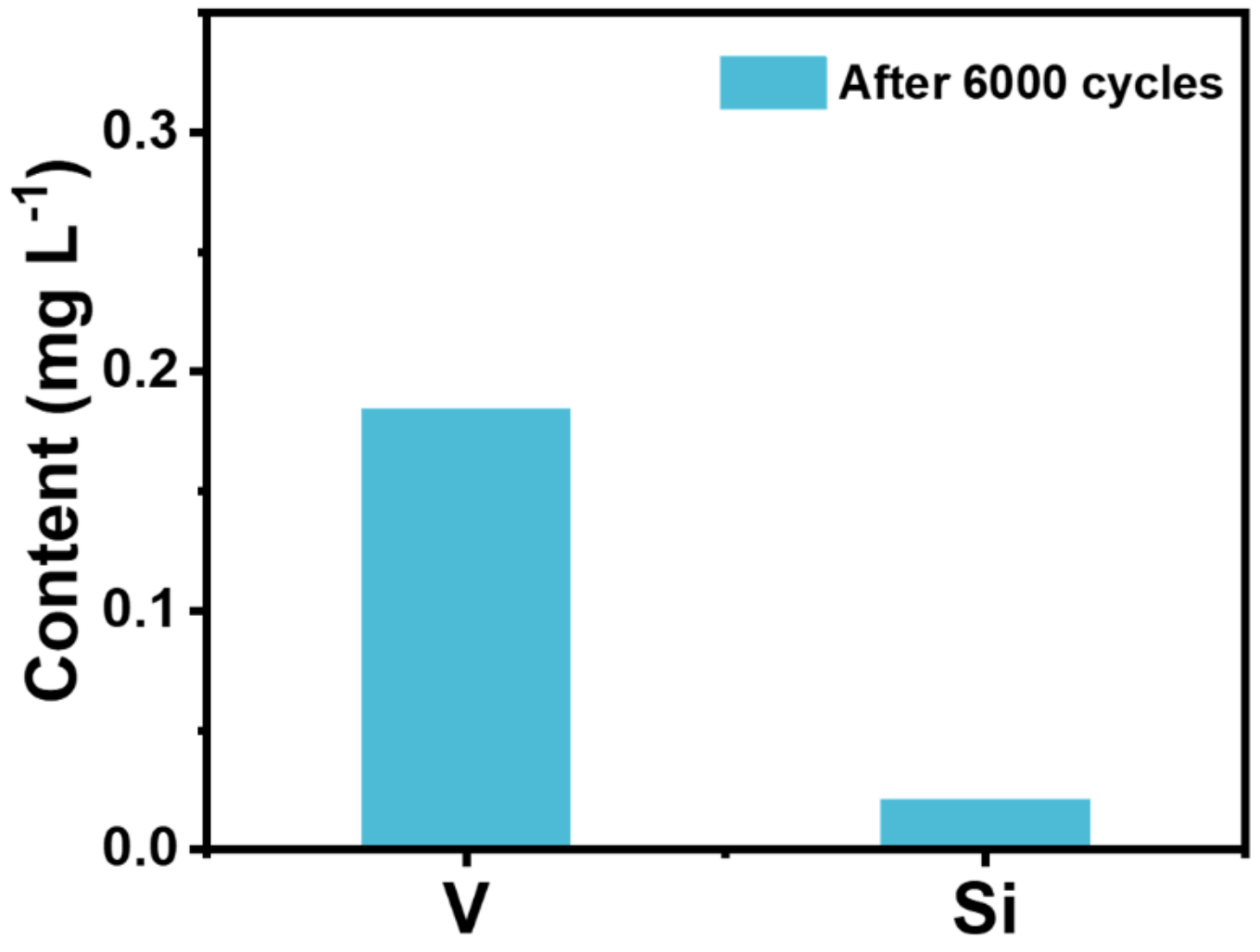

**Figure S8**. Content of V and Si elements of the electrolyte after cycling 6000 CV cycles and cycling test obtained by ICP measurements. And all data are obtained by subtracting the control sample of pure 0.5M $H_2SO_4$.

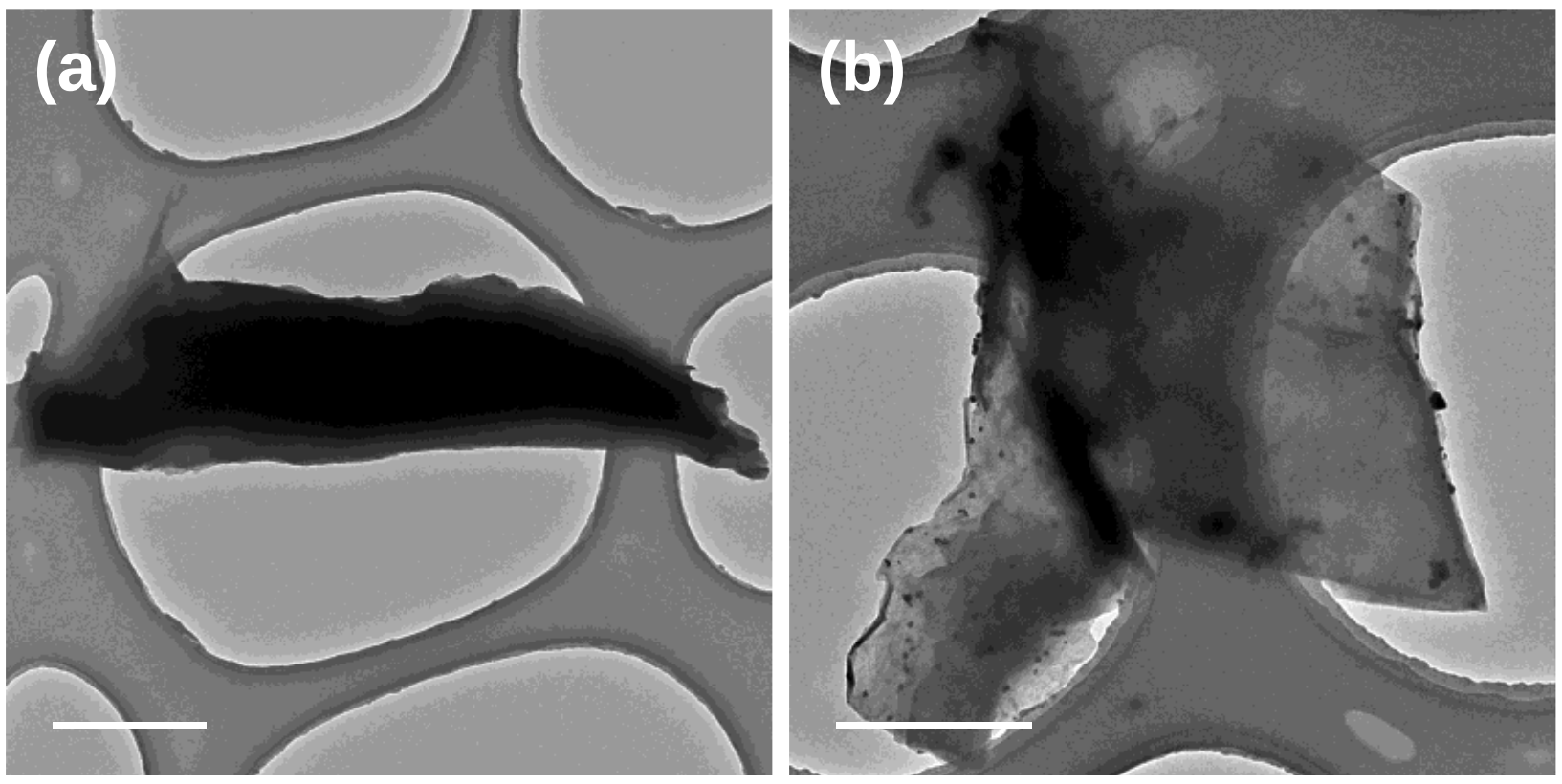

**Figure S9.** TEM images of (a) $V_3Si$ before 6000 CV cycles and (b) $V_3Si$ after 6000 CV cycles.

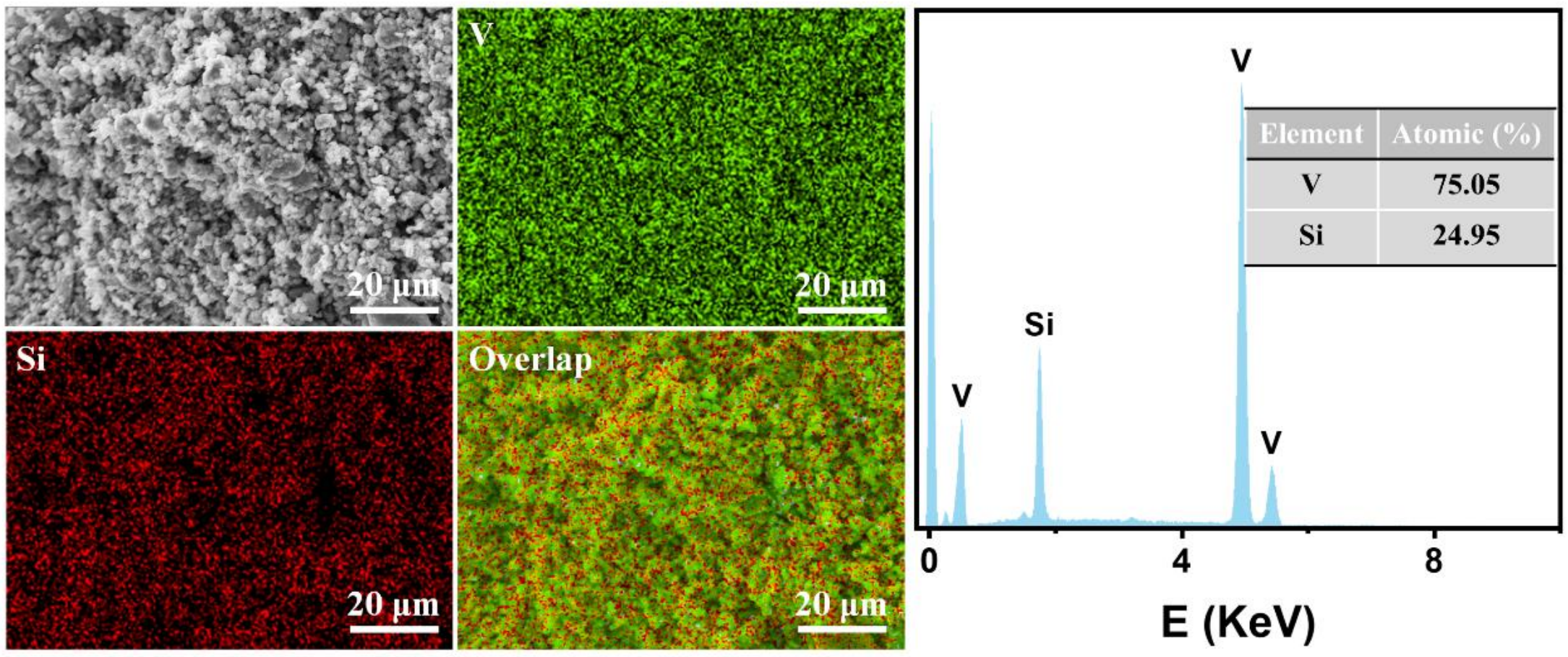

**Figure S10**. SEM image and corresponding EDS mapping images for the $V_3Si$.

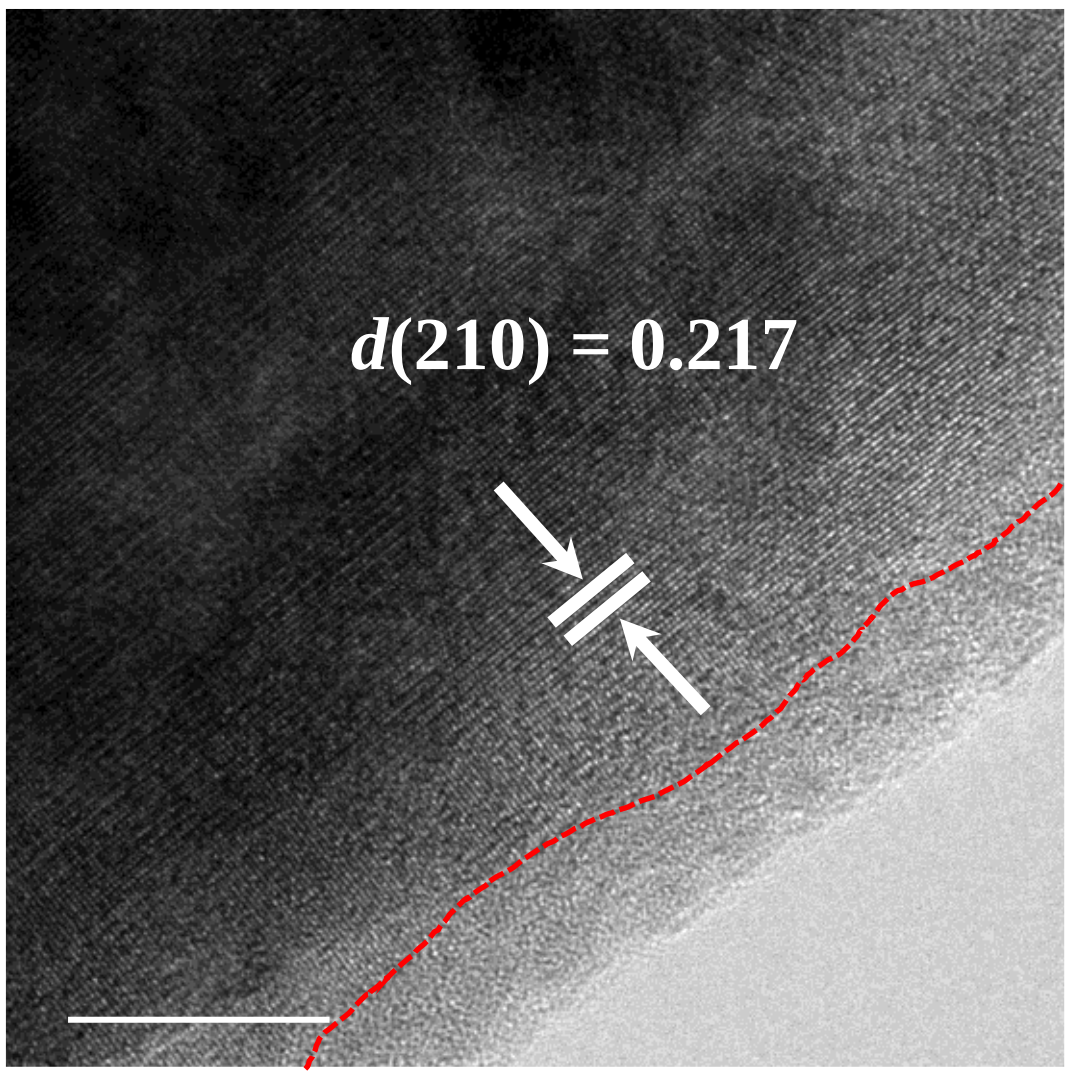

**Figure S11**. HRTEM image of as-synthesized $V_3Si$. The regions outside the red dotted lines indicate that there are amorphous films on the surfaces of the silicide.

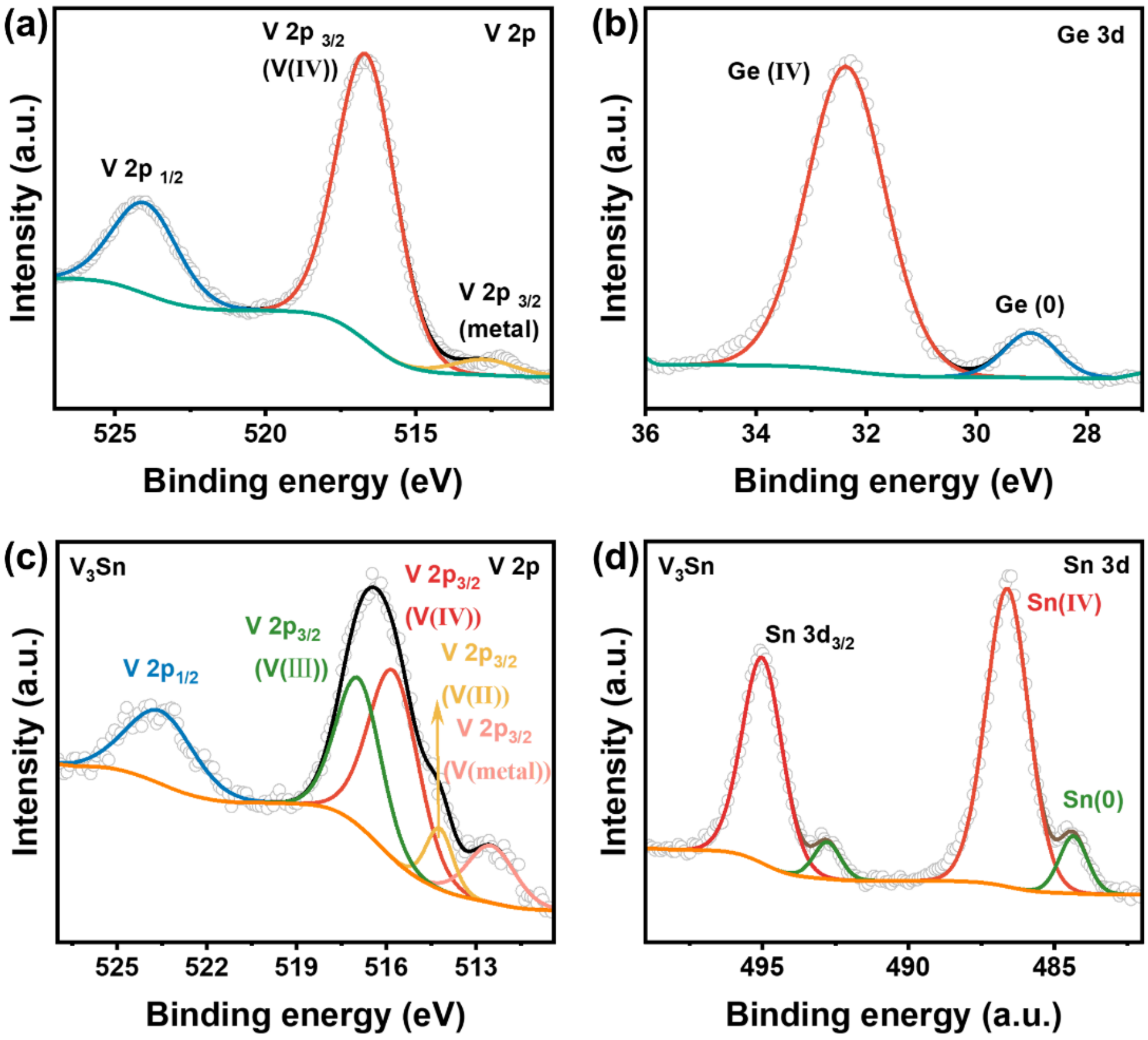

**Figure S12.** High-resolution XPS spectra of V 2p, Ge 3d, and Sn 3d for (a, b) $V_3Ge$ and (c, d) $V_3Sn$.

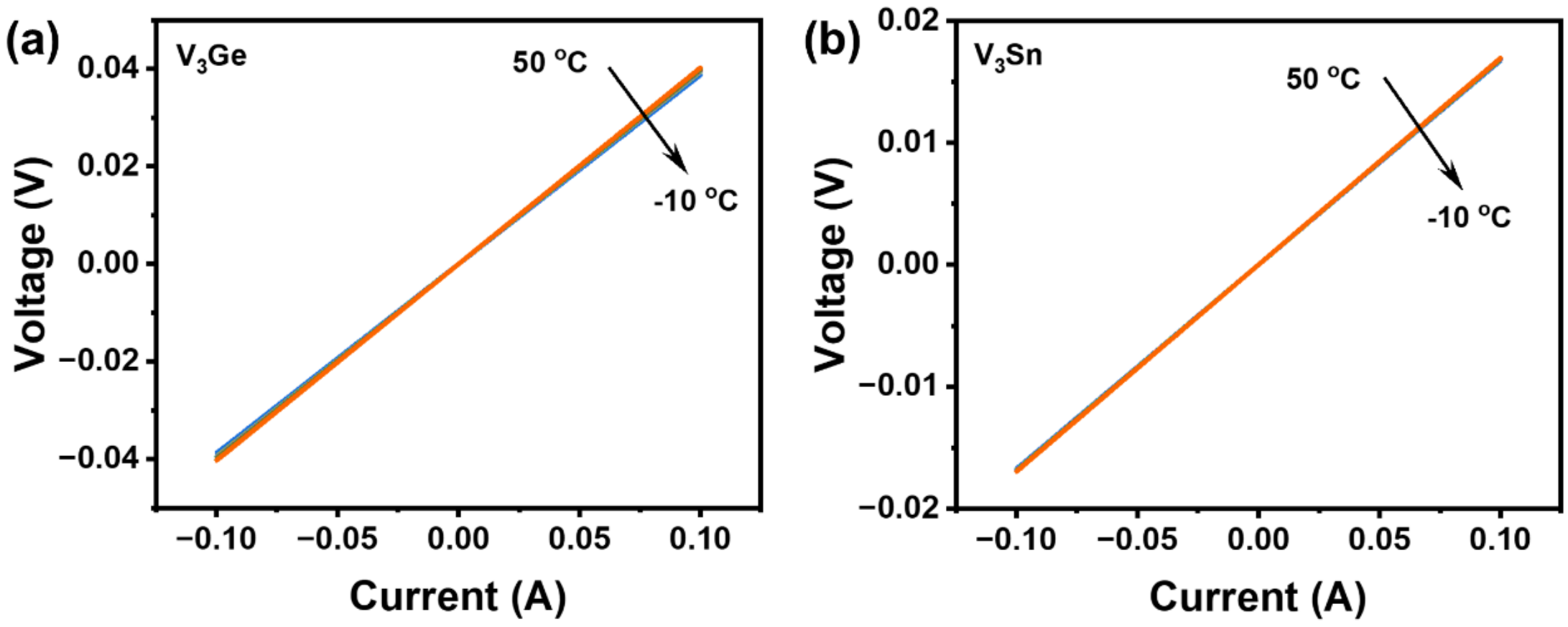

**Figure S13**. I-V curves as a function of temperature of $V_3Ge$ and $V_3Sn$.

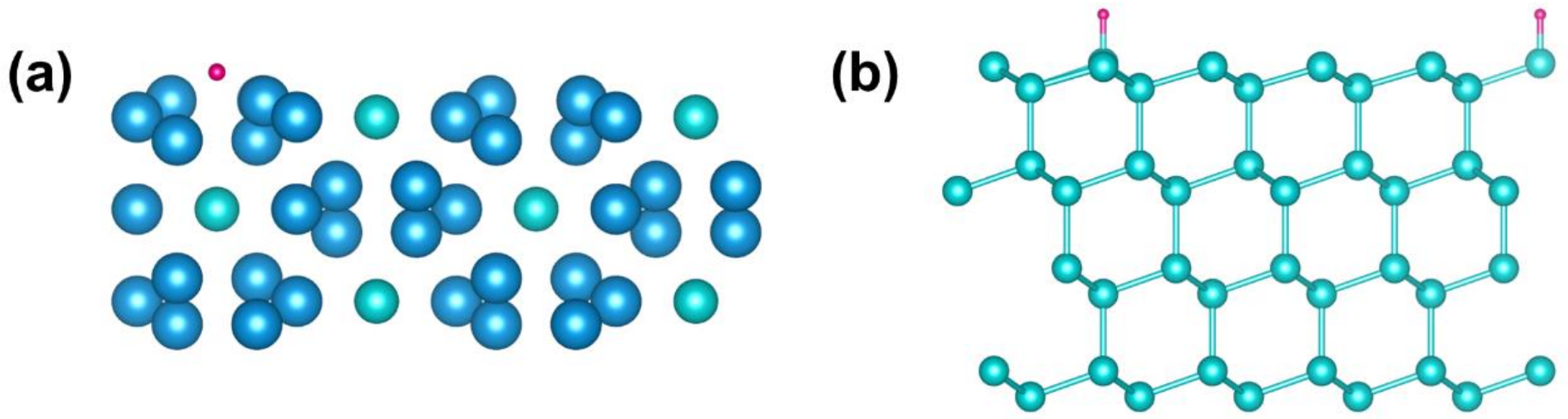

**Figure S14**. Theoretical model used in DFT calculation for (a) $V_3Si$ and (b) Si.

**Table S1.** Summary of the detailed refinement results and crystal structures of intermetallic Si-group compounds.

| Sample | $V_3Si$ | $V_3Ge$ | $V_3Sn$ |
|---|---|---|---|
| PDF card No. | 04-005-8937 | 01-072-1967 | 04-002-9958 |
| Crystal system | Cubic | Cubic | Cubic |
| Space group | *Pm*-3*n* (223) | *Pm*-3*n* (223) | *Pm*-3*n* (223) |
| a = b = c (Å) | 4.7264(2) | 4.78522(16) | 4.9931(14) |
| Volume (Å$^3$) | 105.663492 | 109.573559 | 124.483226 |
| $R_{wp}$ | 4.4% | 5.50% | 5.64% |
| $R_e$ | 2.67% | 1.72% | 1.99% |
| $R_p$ | 2.88% | 3.52% | 3.56% |
| $\chi^2$ | 2.72 | 10.2495 | 8.0127 |

**Table S2.** Structure parameters of intermetallic Si-group compounds.

| | Sample | $V_3Si$ | $V_3Ge$ | $V_3Sn$ |
|---|---|---|---|---|
| V1 | x | 0.25000 | 0.25000 | 0.25000 |
| | y | 0.00000 | 0.00000 | 0.00000 |
| | z | 0.50000 | 0.50000 | 0.50000 |
| | Occ. | 1.000 | 1.000 | 1.000 |
| Si1 | x | 0.00000 | 0.00000 | 0.00000 |
| | y | 0.00000 | 0.00000 | 0.00000 |

| | | | |
|---|---|---|---|
| z | 0.00000 | 0.00000 | 0.00000 |
| Occ. | 1.000 | 1.000 | 1.000 |

**Table S3**. Comparison of the HER activities for $V_3Si$ superconductor and the reported Si-based electrocatalysts.

| **Catalysts** | **Electrolytes** | **Overpotential@j (mV@mA cm$^{-2}$)** | **Tafel slope (mV dec$^{-1}$)** | **Ref.** |
|---|---|---|---|---|
| **$V_3Si$** | **0.5 M $H_2SO_4$** | **33.4@10** | **32.2** | **This work** |
| Rh/Si | 0.5 M $H_2SO_4$ | 110@50 | - | [1] |
| Ir/Si | 0.5 M $H_2SO_4$ | 22@10 | 20 | [2] |
| Os/Si | 0.5 M $H_2SO_4$ | 43@10 | 24 | [3] |
| RuSi | 0.5 M $H_2SO_4$ | 19@10 | 30.2 | [4] |
| $Pd_2Si$ | 0.5 M $H_2SO_4$ | 192@10 | - | [5] |
| PdSi | 0.5 M $H_2SO_4$ | 44@10 | 32.3 | [6] |
| RhSi | 0.5 M $H_2SO_4$ | 114@10 | 50.8 | [6] |
| TiSi | 0.5 M $H_2SO_4$ | 34@10 | 27.1 | [6] |
| RuGe | 0.5 M $H_2SO_4$ | 43@10 | 36.4 | [7] |

**Table S4**. Comparison of the HER activities for $V_3Si$ superconductor and the reported typical electrocatalysts.

| **Catalysts** | **Electrolytes** | **Overpotential@j (mV@mA cm$^{-2}$)** | **Tafel slope (mV dec$^{-1}$)** | **Ref.** |
|---|---|---|---|---|
| **$V_3Si$** | **0.5 M $H_2SO_4$** | **33.4@10** | **32.2** | **This work** |
| Ru@$C_2N$ | 0.5 M $H_2SO_4$ | 22@10 | 30 | [8] |
| RuB | 0.5 M $H_2SO_4$ | 22@10 | 30.7 | [9] |
| Pd/Cu-Pt NRs | 0.5 M $H_2SO_4$ | 22.8@10 | 25 | [10] |
| Pt-$MoO_{3-x}$ | 0.5 M $H_2SO_4$ | 23.3@10 | 28.8 | [11] |
| 400-SWMT/Pt | 0.5 M $H_2SO_4$ | 27@10 | 38 | [12] |
| $PdP_2$@CB | 0.5 M $H_2SO_4$ | 27.5@10 | 29.5 | [13] |
| $RuTe_2$ | 0.5 M $H_2SO_4$ | 33@10 | - | [14] |
| $RuP_2$@NPC | 0.5 M $H_2SO_4$ | 38@10 | 38 | [15] |
| Ru/$C_3N_4$/C | 0.5 M $H_2SO_4$ | ~75@10 | - | [16] |
| Pd-$MoS_2$ | 0.5 M $H_2SO_4$ | 78@10 | 62 | [17] |
| Pt@PCM | 0.5 M $H_2SO_4$ | 105@10 | 65.3 | [18] |
| Pt-$MoS_2$ | 0.1 M $H_2SO_4$ | ~150@10 | 96 | [19] |
| Pt-GDY2 | 0.5 M $H_2SO_4$ | ~50@30 | 38 | [20] |
| IrNi NCs | 0.5 M $H_2SO_4$ | 32@20 | - | [21] |
| NiRu@N-C | 0.5 M $H_2SO_4$ | 50 | 36 | [22] |
| Ni@$Ni_2$P-Ru | 0.5 M $H_2SO_4$ | 51 | 35 | [23] |
| RuPx@NPC | 0.5 M $H_2SO_4$ | 51 | 46 | [24] |
| Ru-$MoO_2$ | 0.5 M $H_2SO_4$ | 55 | 44 | [25] |
| Pt@NHPCP | 0.5 M $H_2SO_4$ | 57 | 27 | [26] |
| Ru-HPC | 0.5 M $H_2SO_4$ | 61.6 | 66.8 | [27] |
| s-$RuS_2$/S-rGO | 0.5 M $H_2SO_4$ | 69 | 64 | [28] |
| Ni-doped $RuO_2$ | 0.5 M $H_2SO_4$ | 78 | - | [29] |
| RuMeOH/THF | 0.5 M $H_2SO_4$ | 83 | 46 | [30] |
| Te@Ru | 0.5 M $H_2SO_4$ | 86 | 36 | [31] |
| Ru@CN | 0.5 M $H_2SO_4$ | 126 | - | [32] |